\documentclass[12pt,a4paper]{article} 
\pdfoutput=1
\usepackage{jheppub}
\usepackage{enumerate}
\usepackage{epsfig} 
\usepackage{float} 
\usepackage[caption = false]{subfig}
\usepackage{wasysym} 
\usepackage{mathrsfs} 
\usepackage{amsfonts} 
\usepackage{amsbsy} 
\usepackage{amscd} 
\usepackage{pstricks} 
\usepackage{multirow} 
\usepackage{tikz}
\usepackage{color}
\usepackage{slashed}
\usepackage{array}
\usepackage{tablefootnote}

\usetikzlibrary{arrows,positioning,shapes.geometric} 
\usepackage{color}
\definecolor{myred}{rgb}{0.6,0,0} 
\definecolor{myblue}{rgb}{0,0.2,0.4}
\definecolor{mygreen}{rgb}{0,0.9,0.1}
\definecolor{hc}{rgb}{.9,0.1,0.7}
\definecolor{hcout}{rgb}{.9,0.7,0.9}
\definecolor{Orange}{rgb}{1.,0.65,0.}

\newcommand{\be}{\begin{equation}}
\newcommand{\ee}{\end{equation}}
\newcommand{\bea}{\begin{eqnarray}}
\newcommand{\eea}{\end{eqnarray}}

\title{Jet substructure shedding light on heavy Majorana neutrinos at the LHC}

\author[a]{Arindam Das,}
\author[b]{Partha Konar,}
\author[c]{and Arun Thalapillil}
\affiliation[a]{School of Physics, KIAS, Seoul 130-722, Korea}
\affiliation[b]{Theoretical Physics Group, Physical Research Laboratory, Ahmedabad-380009, India}
\affiliation[c]{Indian Institute of Science Education and Research,
Homi Bhabha Rd, Pashan, Pune 411 008, India}
\emailAdd{arindam@kias.re.kr}
\emailAdd{konar@prl.res.in}
\emailAdd{thalapillil@iiserpune.ac.in}

\abstract{
The existence of tiny neutrino masses and flavor mixings can be explained naturally in various seesaw models, many of which typically having additional Majorana type SM gauge singlet right handed neutrinos ($N$). If they are at around the electroweak scale and furnished with sizeable mixings with light active neutrinos, they can be produced at high energy colliders, such as the Large Hadron Collider (LHC). A characteristic signature would be same sign lepton pairs, violating lepton number, together with light jets -- $pp\to N\ell^{\pm}, \; N\to\ell^{\pm}W^{\mp}, \; W^{\mp}\to jj$. We propose a new search strategy utilising jet substructure techniques, observing that for a heavy right handed neutrino mass $M_N$ much above $M_{W^\pm}$, the two jets coming out of the boosted $W^\pm$ may be interpreted as a single fat-jet ($J$). Hence, the distinguishing signal topology will be $\ell^{\pm}\ell^{\pm} J$. Performing a comprehensive study of the different signal regions along with complete background analysis, in tandem with detector level simulations, we compute statistical significance limits. We find that heavy neutrinos can be explored effectively for mass ranges $300$ GeV $\leq M_N \leq 800$ GeV and different light-heavy neutrino mixing $|V_{\mu N}|^{2}$. At the 13 TeV LHC with 3000 $\mathrm{fb}^{-1}$ integrated luminosity one can competently explore mixing angles much below present LHC limits, and moreover exceed bounds from electroweak precision data. 
}


\begin{document}

\maketitle

\section{Introduction}
\label{sec:intro}

The experimental evidence for neutrino oscillations \cite{Neut1,Neut2,Neut3,Neut4,Neut5,Neut6} and lepton flavor mixings, from the various experiments, motivate extensions of the SM incorporating non-zero neutrino masses and mixings. 
After the pioneering realization of the unique $d=5$ operator \cite{Weinberg:1979sa} within the SM  with $\Delta L=2$ lepton number violation $(L= \rm{Lepton~number})$, it was realized that the Seesaw mechanism \cite{seesaw0,seesaw1,seesaw2,seesaw3,seesaw4,seesaw5,seesaw6} could be the simplest idea to explain the smallness of the neutrino masses and flavor mixings. In many of these models, SM is extended by gauge singlet, Majorana type, heavy right handed neutrinos (RHNs).  After electroweak (EW) symmetry breaking, the light Majorana neutrino masses are generated by, for instance, the so called type-I seesaw mechanism. 

Through the seesaw mechanism, the flavor eigenstates of the SM light neutrino mix with the mass eigenstates of the light neutrinos and RHNs. The SM singlet RHNs ($N$) interact with the SM gauge bosons through lepton mixing. Such Majorana type RHNs, if at the EW scale, can be produced at the Large Hadron Collider (LHC) with a distinguishing signature -- Same Sign Di-Leptons (SSDL) and di-jets. In this channel the heavy RHNs decay into a $W^{\pm}$ and a lepton. In cases where the RHNs are sufficiently massive, very often the gauge bosons are significantly boosted, resulting in collimated energy deposits in the hadronic calorimeter. With a suitable jet algorithm, these collimated hadron four momenta may be reconstructed as a fat-jet $(J)$. Fat-jets retain information of their origins and have several distinct properties that may be leveraged for tagging and signal discrimination. The resulting signal of interest therefore becomes SSDL + fat-jet. In this paper we consider searches for RHNs with masses $M_N \geq 300$ GeV, which is sufficient to produce boosted jets. It is important to search for such relatively small-mass RHNs at colliders, since from a very general theoretical viewpoint, a small $M_N$ may be considered technically natural~\cite{{Fujikawa:2004jy}, {deGouvea:2005er}}. This is because in the limit $M_N \rightarrow 0$ one regains $U(1)_{\text{\tiny{B-L}}}$ as a global symmetry of the Lagrangian. RHN phenomenologies in $U(1)$ extended models have been studied in \cite{Kang:2015uoc, Accomando:2016rpc, Accomando:2017qcs}. Interesting phenomenological aspects of the RHN in the Left-Right (LR) model have been studied in \cite{Gluza:2015goa, Lindner:2016lxq, Das:2017hmg}. Different experiments such as ATLAS \cite{Aad:2015xaa} and CMS \cite{CMS8:2016olu, Khachatryan:2015gha} have already searched for RHNs in the SSDL + dijets channel, assuming non-boosted $W^\pm$. 

At the 8 TeV LHC, with $20.3$ fb$^{-1}$ luminosity and $95\%$ confidence limit (C.~L.), ATLAS \cite{Aad:2015xaa} has probed mixings for muon flavor down to a $|V_{\mu N}|^2$ of $3.5\times 10^{-3}$, for $M_N= 100$ GeV. The limits further goes down to $2.9\times 10^{-3}$ for $M_N=110$ GeV and then monotonically weakens with mass, up to $M_N=500$ GeV. At  $M_N=500$ GeV the limits are $|V_{\mu N}|^2=4\times10^{-1}$. The limits are nearly two orders of magnitude weaker in the case of electron flavor mixings $|V_{e N}|^2$ at the $95\%$ C.~L. 

CMS has also studied the SSDL plus dijet signal and obtain the exclusion limits for $|V_{eN}|^2$ \cite{CMS8:2016olu} and $|V_{\mu N}|^2$ \cite{Khachatryan:2015gha}. Both studies are performed at the 8 TeV LHC with $19.7$ fb$^{-1}$ luminosity at  $95\%$ C.~L. The limits for the mixed $e^{\pm}\mu^{\pm}+jj$ final state was also considered in \cite{CMS8:2016olu}. CMS observed upper limits for $|V_{eN}|^2$ at $1.2\times10^{-4}$ for $M_N=40$ GeV, $2\times 10^{-2}$ for $M_N=85$ GeV, $8\times10^{-3}$ for $M_N=130$ GeV and $1.2\times10^{-2}$ for $M_N=200$ GeV. Thus, the $|V_{eN}|^2$ limits were found to again weaken with $M_N$. Alternatively, RHNs may be excluded as large as $M_N=480$ GeV, assuming the mixing is unity. The limits on $|V_{\mu N}|^2$ from the SSDL + dijet final state with $\mu$ flavor  is probed down to $2\times  10^{-5}$ for $M_N=40$ GeV, $4.5\times 10^{-3}$ for $M_N=90$ GeV, $1.75\times 10^{-3}$ for $M_N=125$ GeV and $7\times 10^{-3}$ for $M_N=175$ GeV with $|V_{\mu N}|^2$ again weakening subsequently with $M_N$. For $M_N=500$ GeV the limit is $|V_{\mu N}|^2=0.6$. 

In this paper we leverage boosted $W^{\pm}$ production from massive RHN, and its subsequent decay into a fat-jet in association with $\mu^\pm \mu^\pm$ pairs. The $P_T$ of the $W^\pm$ scale as $P_T^W \sim (M_N^2-M_{W}^2)/M_N$ and the separation between the hadronic decay products of $W^\pm$ scale as $\sim M_W/P_T^W$. Therefore, a natural region of focus may be the intermediate to heavy RHN mass range, say $M_N\geq 300$ GeV. In this mass range, the only other competent limit that exists comes from indirect EW precision data (EWPD). The EWPD limit is around $|V_{\mu N}|^{2} =0.009$~\cite{deBlas:2013gla, delAguila:2008pw,Akhmedov:2013hec}. The mixing limits may also be obtained from the Higgs data \cite{BhupalDev:2012zg,Das:2017zjc, Das:2017rsu} for  $10$ GeV $\leq M_N \leq 200$ GeV.

\begin{figure}[t]
\begin{center}
\includegraphics[bb=0 0 581 214, scale=.5]{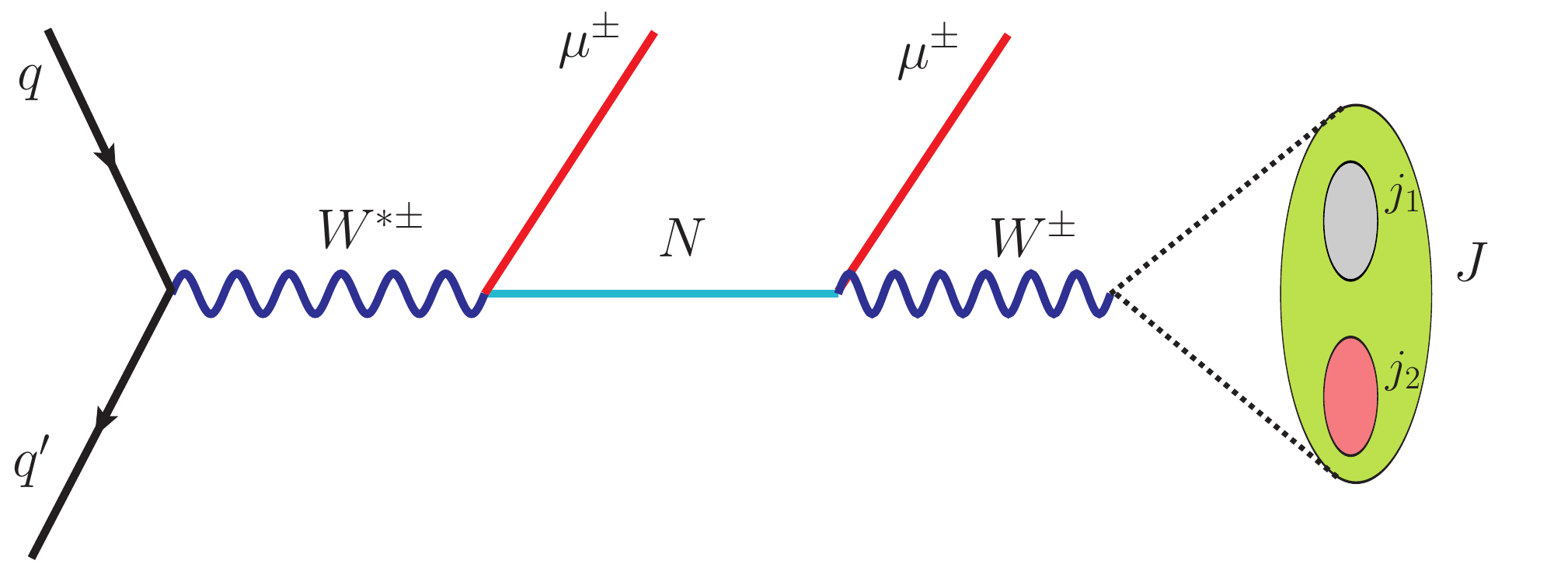} 
\end{center}
\caption{
SSDL + fat-jet production at the LHC.}
 \label{fig:feyn}
\end{figure}

For simplicity and clarity, we consider only the $\mu$ flavor for the SSDL. Moreover, $\mu$ detection efficiencies are better, compared to electrons and tau leptons. We place limits on $|V_{\mu N}|^2$ at the 13 TeV LHC, with 3000 fb $^{-1}$ luminosity, in the $300$ GeV $\leq M_N \leq 800$ GeV mass range. A representative diagram for the parton level production and decay of RHN, leading to final states of interest, is shown in figure~\ref{fig:feyn}.

Search strategies utilising boosted and collimated objects have proven to be spectacularly successful in searches at the LHC. The seminal ideas \cite{Seymour:1993mx, Butterworth:2007ke, Brooijmans:1077731,Butterworth:2008iy} have burgeoned into many sophisticated methods that enable tagging jets arising from the decay of boosted heavy particles,  improving searches for new topologies, investigating jet properties and mitigating  underlying events and pile-up (please see \cite{Adams:2015hiv} and references therein for a review of some these techniques).

In the context of sterile neutrinos and related models there have been a few studies that have, in a broader sense, leveraged the effectiveness of collimated objects in the signal topology \cite{Izaguirre:2015pga,Antusch:2016ejd,Mitra:2016kov,Dube:2017jgo,Cox:2017eme}. Nevertheless, surprisingly, there have not been any investigations in the SSDL+fat-jet channel, in the RHN collider search context. We utilise for the first time, jet substructure techniques to augment RHN searches, in the $l^\pm l^\pm J$ channel corresponding to figure~\ref{fig:feyn}.

The paper is organized as follows. In section~\ref{sec:model}, we discuss the prototypical model along with the RHN production cross sections at the 13 TeV LHC. We also calculate the decay widths and the corresponding branching ratios there. In section~\ref{sec:fat}, we briefly describe the fat-jet technique for W-tagging.
Sections~\ref{sec:calc} and \ref{sec:res} are dedicated to the setup, collider analysis, discussion of kinematic distributions, and presentation of the salient results and limits. We conclude in section~\ref{sec:conc}.

\section{Model and heavy Majorana neutrinos at the LHC}
\label{sec:model}
In the simplest model of seesaw, we only introduce SM gauge-singlet Majorana RHNs $N_R^{\beta}$ (where $\beta$ is a flavor index). $N_R^{\beta}$ would couple with the SM lepton doublet $\ell_{L}^{\alpha}$ and the Higgs doublet $H$.
The relevant part of the Lagrangian density is
\bea
\mathcal{L} \supset -Y_D^{\alpha\beta} \overline{\ell_L^{\alpha}}H N_R^{\beta} 
                   -\frac{1}{2} M_N^{\alpha \beta} \overline{N_R^{\alpha C}} N_R^{\beta}  + H. c. .
\label{typeI}
\eea
After EW symmetry breaking by a vacuum expectation value (VEV) $ H =\begin{pmatrix} \frac{v}{\sqrt{2}} ~  0 \end{pmatrix}^{T}$, we obtain the Dirac mass matrix $M_{D}= \frac{Y_D v}{\sqrt{2}}$. Using these Dirac and Majorana mass matrices, we can write the full neutrino mass matrix as 
\bea
M_{\nu}=\begin{pmatrix}
0&&M_{D}\\
M_{D}^{T}&&M_{N}
\end{pmatrix}.
\label{typeInu}
\eea
Diagonalizing this matrix, we obtain the well-known seesaw formula for the light Majorana neutrinos 
\bea
m_{\nu} \simeq - M_{D} M_{N}^{-1} M_{D}^{T}.
\label{seesawI}
\eea
With $M_{N}\sim 100$ GeV, we require $Y_{D} \sim 10^{-6}$  for $m_{\nu}\sim 0.1$ eV. However, in the general parameterization for the seesaw formula \cite{Casas:2001sr}, $Y_{D}$ can be large and sizable, which is the case we are going to consider in this paper. An interesting class of models have mass matrices $M_D$ and $M_N$ with specific textures, enforced by some symmetries \cite{Pilaftsis:2003gt, Kersten:2007vk, Xing:2009in, He:2009ua, Ibarra:2010xw, Deppisch:2010fr,Dev:2013oxa}, so that a large light-heavy neutrino mixing occurs even at a low scale, satisfying the neutrino oscillation data. 

If these RHNs reside at the electroweak scale, then they can be produced in high energy colliders such as the LHC with a variety of phenomenological aspects\cite{Keung:1983uu,Datta:1992qw, Datta:1993nm, AguilarSaavedra:2009ik, Chen:2011hc, Das:2012ze, Das:2014jxa, Das:2015toa, Dev:2015pga, Das:2016akd, Gluza:2016qqv, delAguila:2008hw, delAguila:2007qnc, AguilarSaavedra:2012gf, Nayak:2015zka, Nayak:2013dza, AguilarSaavedra:2012fu, LalAwasthi:2011aa, Fong:2011xh, Dias:2011sq, Ibarra:2011xn, ILC1, ILC2, type-I1, type-I2, Batell:2015aha, Leonardi:2015qna, p-photon, konar1, konar2,  Dutta:1994wz, Haba:2009sd, Matsumoto:2010zg, Mondal:2012jv, Helo:2013esa, Hessler:2014ssa, Deppisch:2015qwa, Arganda:2015ija, Dib:2015oka,Campos:2017odj, Dev:2015kca, Chen:2013fna, Dev:2017dui}. Searches for Majorana RHNs can be performed via the `smoking-gun'  tri-lepton, as well as, SSDL+dijet signals. The rates will generally be suppressed by the square of light-heavy mixing $|V_{\ell N}|^2\simeq | M_D M_N^{-1}|^2$. 
A comprehensive, general study\footnote{The study uses data from neutrino oscillation experiments \cite{Neut1,Neut2,Neut3,Neut4,Neut5,Neut6}, bounds from Lepton Flavor Violation (LFV) \cite{Adam:2011ch, Aubert:2009ag, OLeary:2010hau}, Large Electron-Positron (LEP) \cite{Achard:2001qv, delAguila:2008pw, Akhmedov:2013hec} experiments using the non-unitarity effects \cite{Antusch:2006vwa, Abada:2007ux} applying the Casas- Ibarra conjecture \cite{Casas:2001sr, Ibarra:2010xw, Ibarra:2011xn, Dinh:2012bp}.} 
 of $|V_{\ell N}|^{2}$ and associated parameters is given in \cite{Das:2017nvm}. 
Bounds may be placed on the light-heavy mixing angles using results from different experiments, as in \cite{Aad:2015xaa, CMS8:2016olu,Khachatryan:2015gha,delAguila:2008pw, Akhmedov:2013hec, deBlas:2013gla, BhupalDev:2012zg, KamLAND-Zen:2016pfg, Das:2017zjc,Das:2017rsu,Achard:2001qv, Rasmussen:2016njh}, considering degenerate Majorana RHNs.

Through the seesaw mechanism, a flavor eigenstate ($\nu$) of the SM neutrino may be expressed in terms of the mass eigenstates of the light ($\nu_m$) and heavy ($N_m$) Majorana neutrinos as 
\bea 
  \nu \simeq  \mathcal{N} \nu_m  + \mathcal{R} N_m \; .
\eea 
Here 
\bea
\mathcal{R}= M_D M_N^{-1}~~,~~~\mathcal{N}=\big(1-\frac{1}{2}\epsilon\big) U_{\rm{PMNS}}\; ,
\eea
with $\epsilon=\mathcal{R^\ast}\mathcal{R}^T$ and $U_{\rm{PMNS}}$ \cite{Pontecorvo:1957qd, Maki:1962mu} the usual neutrino mixing matrix. In terms of mass eigenstates, the charged current interactions for the heavy neutrinos is then given by 
\bea 
\mathcal{L}_{CC} =
 -\frac{g}{\sqrt{2}} W_{\mu}
  \bar{e} \gamma^{\mu} P_L  \Big( \mathcal{N} \nu_m  + \mathcal{R} N_m\Big)  + h.c., 
\label{CC}
\eea
where $e$ denotes three generations of charged leptons, in vector form, and 
  $P_L =\frac{1}{2} (1- \gamma_5)$. Similarly, the neutral current interactiona are given by 
\bea 
\mathcal{L}_{NC} &=& -\frac{g}{2 c_w}  Z_{\mu} 
\Big[ 
  \overline{\nu_m} \gamma^{\mu} P_L ({\cal N}^\dagger {\cal N}) \nu_m 
 +  \overline{N_m} \gamma^{\mu} P_L ({\cal R}^\dagger {\cal R}) N_m  \nonumber \\
&+& \Big\{ 
  \overline{\nu_m} \gamma^{\mu} P_L ({\cal N}^\dagger  {\cal R}) N_m 
  + h.c. \Big\} 
\Big]  , 
\label{NC}
\eea
 where $c_w=\cos \theta_w$ with $\theta_w$ being the weak mixing angle.
 
\begin{figure}[tb]
\begin{center}
\includegraphics[bb=0 0 525 323,scale=0.39]{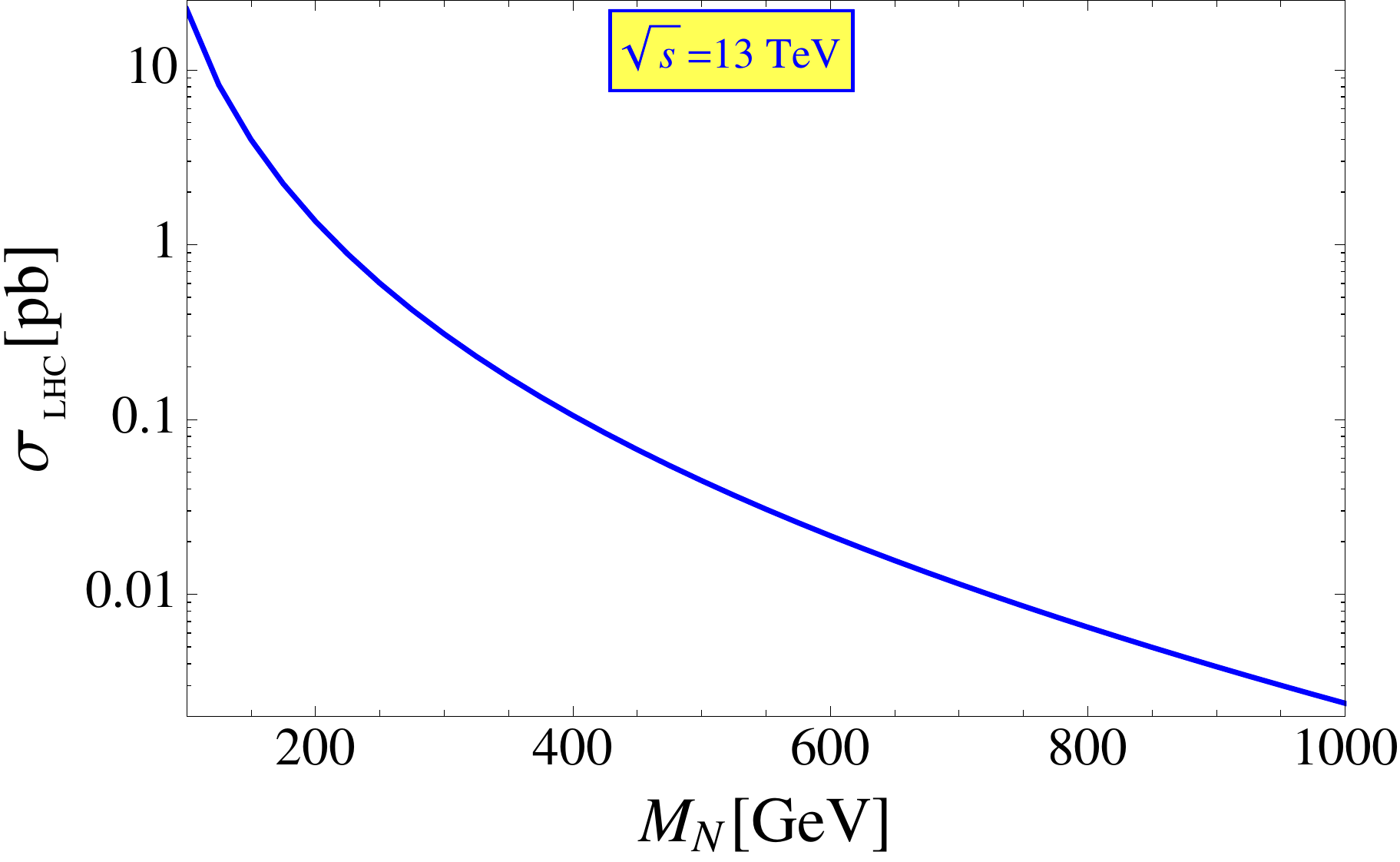} \; \;
\includegraphics[bb=0 0 975 595,scale=0.21]{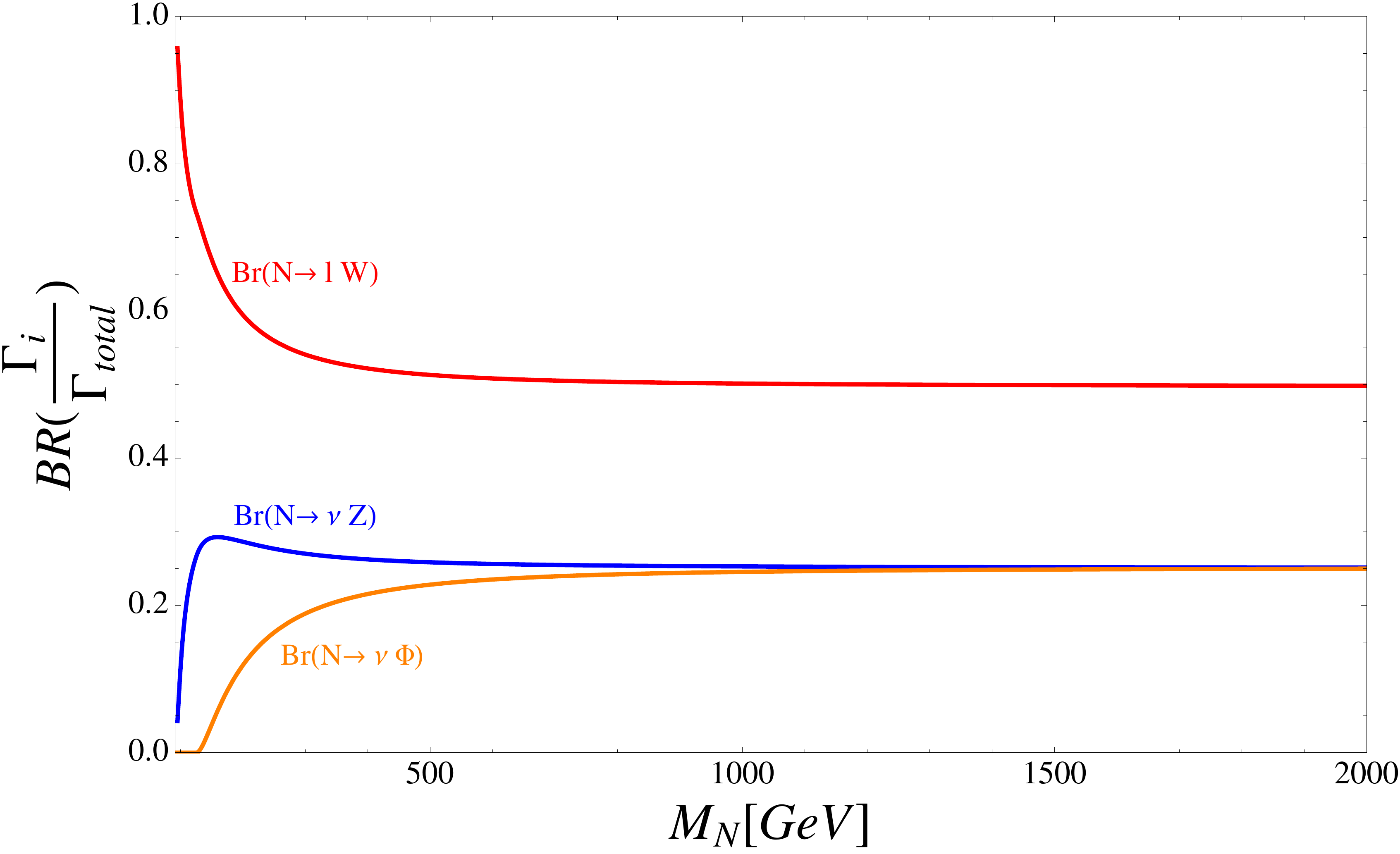}
\end{center}
\caption{
(Left) The total production cross section of the heavy Majorana neutrino 
 as a function of its mass at the LHC with $\sqrt{s}=13$ TeV and normalised by the $|V_{\mu N}|^2$. 
 (Right) Heavy neutrino branching ratios ($BR_i$) for different decay modes as a function of its mass.
 }
 \label{fig:LHC_heavyN}
\end{figure}

At the LHC, the heavy neutrinos can be produced through charged current interactions, via the $s$-channel exchange of W bosons. The main production process at the parton level is $u \bar{d}\rightarrow \mu^{+} N$ (and $\bar{u} d \rightarrow \mu^{-} N$). The differential cross section is found to be 
\bea
\frac{d \hat{\sigma}_{LHC}}{d \cos\theta}
&=&
(3.89 \times 10^8 \;{\rm pb}) \times 
 \frac{\beta}{32 \pi \hat{s}} \frac{\hat{s} +M_N^2}{\hat s}
 \Big( \frac{1}{2} \Big)^2 3 \Big( \frac{1}{3} \Big)^2 
\frac{g^4}{4}  \nonumber \\
&&\frac{({\hat s}^2-M_N^4)(2 + \beta \cos^2 \theta)}
{({\hat s} - M_W^2)^2+ M_W^2 \Gamma_W^2}, 
\eea
 where $\sqrt{\hat s}$ is the center-of-mass energy of the colliding partons, $M_N$ the mass of $N$, and $\beta =({\hat s}-M_N^2)/({\hat s}+M_N^2)$. The total production cross section at the LHC is thus given by 
\bea 
\sigma_{LHC} &=&
\int d \sqrt{\hat s} \int d \cos \theta 
\int^1_{{\hat s}/E_{CMS}^2} dx 
\frac{\sqrt{4 {\hat s}}}{x E_{CMS}^2} 
f_u(x,Q) f_{\bar d}\left( \frac{\hat s}{x E_{CMS}},Q \right)  
\frac{d \hat{\sigma}_{LHC}}{d \cos\theta} \nonumber \\
&+& 
(u \to {\bar u}, {\bar d} \to d) \; .
\label{XLHC}
\eea 
We take $E_{CMS} = 13$ TeV, for the center-of-mass energy of the LHC. In the numerical analysis, we further employ CTEQ5M \cite{Pumplin:2002vw} for the $u$-quark ($f_u$) and ${\bar d}$-quark ($f_{\bar d}$) parton distribution functions, with a factorization scale $Q = \sqrt{\hat s}$. The total cross section thus computed, as a function of $M_N$, is depicted in figure~\ref{fig:LHC_heavyN} (Left pane), normalized by $|V_{\mu N}|^2$. Hence, the resultant cross sections shown in figure~\ref{fig:LHC_heavyN} correspond to maximum values for a fixed $M_N$. 
 
 The main decay modes of the heavy neutrino are 
 $N \to \ell W$, $\nu_{\ell} Z$, $\nu_{\ell} h$. 
The corresponding partial decay widths \cite{Das:2012ze, Das:2014jxa, Das:2015toa, Das:2016hof, Das:2017pvt} are given by
\bea
\Gamma(N \rightarrow \ell W) 
 &=& \frac{g^2 |V_{\ell N}|^{2}}{64 \pi} 
 \frac{ (M_{N}^2 - M_W^2)^2 (M_{N}^2+2 M_W^2)}{M_{N}^3 M_W^2} ,
\nonumber \\
\Gamma(N \rightarrow \nu_\ell Z) 
 &=& \frac{g^2 |V_{\ell N}|^{2}}{128 \pi c_w^2} 
 \frac{ (M_{N}^2 - M_Z^2)^2 (M_{N}^2+2 M_Z^2)}{M_{N}^3 M_Z^2} ,
\nonumber \\
\Gamma(N \rightarrow \nu_\ell h) 
 &=& \frac{ |V_{\ell N}|^2 (M_{N}^2-M_h^2)^2}{32 \pi M_{N}} 
 \left( \frac{1}{v }\right)^2.
\label{widths}
\eea 
Note that the decay width of heavy neutrinos into $W^\pm$ is about twice as large as that into $Z^0$, owing to the two degrees of freedom. 
We plot the branching ratios $BR_i \left(\equiv{\Gamma_{i}}/{\Gamma_{\rm total}}\right)$ of the various decay modes $\left(\Gamma_{i}\right)$ in figure~\ref{fig:LHC_heavyN} (Right pane). Note that for larger values of $M_{N}$, the branching ratios are related as
\bea
BR\left(N\rightarrow \ell W\right) : BR\left(N\rightarrow \nu Z\right) : BR\left(N\rightarrow \nu H\right) \simeq 2: 1: 1.
\eea

As mentioned earlier, in our analysis we will consider Majorana RHNs having mass in the range $300$ GeV $\le M_N \le 800$ GeV. In this mass range, the $W^\pm$ boson from the leading decay mode $N\rightarrow\ell W$ (see figure~\ref{fig:LHC_heavyN}), will be boosted. These boosted $W^\pm$ can decay hadronically to produce a fat-jet, with the characteristic final state $\mu^{\pm}\mu^{\pm}J$.

\section{Fat-jets and Jet Substructure for $W$-like jet tagging} 
\label{sec:fat}
As we have emphasized, in scenarios where the right-handed neutrino is very heavy, the hadronically decaying daughter $W^\pm$ will typically have a large boost. This causes the jets from the $W^\pm$ to be very collimated and one would detect them as a single jet -- a `fat-jet' ($J$). The boosted topology and its associated substructure is extremely powerful in reducing backgrounds, mitigating underlying event contamination and in event tagging \cite{Adams:2015hiv}. In our context, the jet substructure analysis primarily appears as a means to efficiently tag the hadronically decaying boosted-$W^\pm$. Our strategy will be to leverage two variables --  N-subjettiness and jet-mass -- to achieve efficient W-tagging in the $\mu^\pm \mu^\pm J$ final state. 

N-subjettiness \cite{Thaler:2010tr,Thaler:2011gf} is an inclusive jet shape variable defined as 
\begin{equation}
 \tau_N = \frac{1}{\mathcal{N}_0} \sum\limits_i p_{i,T} \min \left\lbrace \Delta R _{i1}, \Delta R _{i2}, \cdots, \Delta R _{iN} \right\rbrace.
 \label{eq:nsub_N}
\end{equation}
The normalization is defined as $\mathcal{N}_0=\sum\limits_i p_{i,T} R$. $i$ runs over the constituent particles in the jet. $p_{i,T}$ are  transverse momenta of the constituent particles, $ \Delta R_{i\alpha} = \sqrt{(\Delta \eta)^2_{i\alpha}+(\Delta \phi)^2_{i\alpha}}$  is the $\eta-\phi$ distance between a candidate $\alpha$-subjet and a constituent particle $i$ and $R$ is the jet radius. $\tau_N$ tries to quantify if the original jet consists of N daughter subjets. A low value of $\tau_N$ suggests that the original jet consists of $N$ or fewer daughter subjets. Thus, information from $\tau_N$ may potentially be used to identify an object that has an N-prong hadronic decay. In fact, it has been shown that a better discriminant to tag an N-subjet object is to consider ratios $\tau_N/\tau_{N-1}$ \cite{Thaler:2010tr,Thaler:2011gf}. For W-tagging, the $W^\pm$ yields two subjets that are collimated, and hence the variable of interest is $\tau_{21}^J=\tau_2/\tau_1$. The mass of the fat-jet ($M_J$), after suitable jet grooming, is another variable that can help in distinguishing signal events from background. At each iteration in a sequential recombination jet algorithm, in the E-scheme, the mother proto jet four-momentum is the vector sum of the daughter proto jet four-momenta. In this fashion, the jet algorithm at the end of the iteration provides a $P_T^J$ for the full fat-jet. $M_J^2$ is computed as the invariant mass square of the fat-jet four momentum ($P_J^2$).

To reconstruct the candidate fat-jet, \texttt{Delphes 3.3.2} \cite{deFavereau:2013fsa} hadron calorimeter outputs are clustered using \texttt{FastJet 3.1.3} \cite{{hep-ph/0512210},{Cacciari:2011ma}}. The $\tau_{21}^J$ is computed with the aid of the N-subjettiness extension, available as part of the \texttt{FastJet-contrib} \cite{{hep-ph/0512210},{Cacciari:2011ma}}. Following \cite{Khachatryan:2014vla} for W-tagging, we will choose for the jet clustering algorithm Cambridge-Achen \cite{Dokshitzer:1997in,Wobisch:1998wt} with a jet-cone radius $R=0.8$. We will in addition require specific cuts on $\tau_{21}^J$ and $M_J$ for efficient W-tagging, as we shall discuss in section~\ref{sec:res}.
\section{Analysis setup and Simulation}
\label{sec:calc}

In preparation for our exploration of the SSDL+fat-jet channel at 13 TeV LHC, along with establishing the setup in terms of signal RHN model files and a jet substructure analysis strategy, we must also consider the relevant backgrounds carefully. Towards this end we will perform detailed background simulations and study  the prospects of our proposed channel, in terms of statistical significance. 

Consider the production of heavy RHN, through an off-shell $W^\pm$. This in a further decay produces relatively clean, same sign di-muon pair $\mu^\pm \mu^\pm$ final states, in association with a boosted $W^\pm$. Our primary objective is to unmask these $W^\pm$ from other hadronic backgrounds. This is efficiently achieved by utilizing jet substructure to W-tag the fat-jet originating from $W^\pm$. Of course, one expects from our previous discussions that the fat-jet and jet substructure techniques become significant when $W^\pm$ bosons are generated with sufficient boost. Hence, as mentioned earlier, our primary region of interest is when $M_N \gg M_W$. Notably, these are also the ranges where conventional searches at colliders fail to probe the mixing parameters very effectively. Corresponding to the signal production channel depicted in figure~\ref{fig:feyn}, we will consider
\bea
pp && \rightarrow \ell_{1}^{+} N, \; \; N \rightarrow \ell_{2}^{+} W^{-}, \; \;  W^{-} \rightarrow   J   \nonumber \\
pp  && \rightarrow \ell_{1}^{-} \overline{N}, \; \;  \overline{N} \rightarrow \ell_{2}^{-} W^{+},  \; \; W^{+} \rightarrow J. 
\label{eq:trilep}
\eea
As mentioned before, for concreteness we assume the light-heavy mixing is non-zero only for the muon flavor in a simplified model. The muons also provide cleaner lepton signals. Hence, all leptons we consider in this study will be muons. It is straightforward to extend the analysis if more lepton flavors are allowed. 

Backgrounds for our SSDL+fat-jet channel can originate from electroweak gauge boson decays along with a fat-jet; the latter for instance produced from a W boson decaying to $J$. Additionally, some of the QCD jets can also mimic $J$.  Hence, one is required to simulate all such processes accompanied by hard jet(s) at the parton level, and then match them with shower jet events.

Dominant contributions come from same-sign $W^\pm$ pair production in association with jets -- $W^\pm W^\pm + jets$. Here,  $W^\pm$ would decay leptonically. One of these jets has the possibility to resemble a $W^\pm$-like fat-jet.  Another significantly large contribution comes from $WZ$ production, where both vector bosons decay leptonically. Subsequently, one of the charged lepton is missed in the detector, giving an SSDL signature. An additional fat-jet like component can come either from a radiated jet or an associated $W^\pm$ boson decaying hadronically.  As demonstrated in \cite{Khachatryan:2015gha} backgrounds from the top quark decays can be controlled effectively by rejecting the events where at least one jet had been identified as originating from the b-quark. Additional veto affects our signal and other backgrounds at 5-7\% level \cite{Bhardwaj:2018lma}.

\begin{figure}[t!]
\begin{center}
\includegraphics[bb=0 0 548 323,scale=0.6]{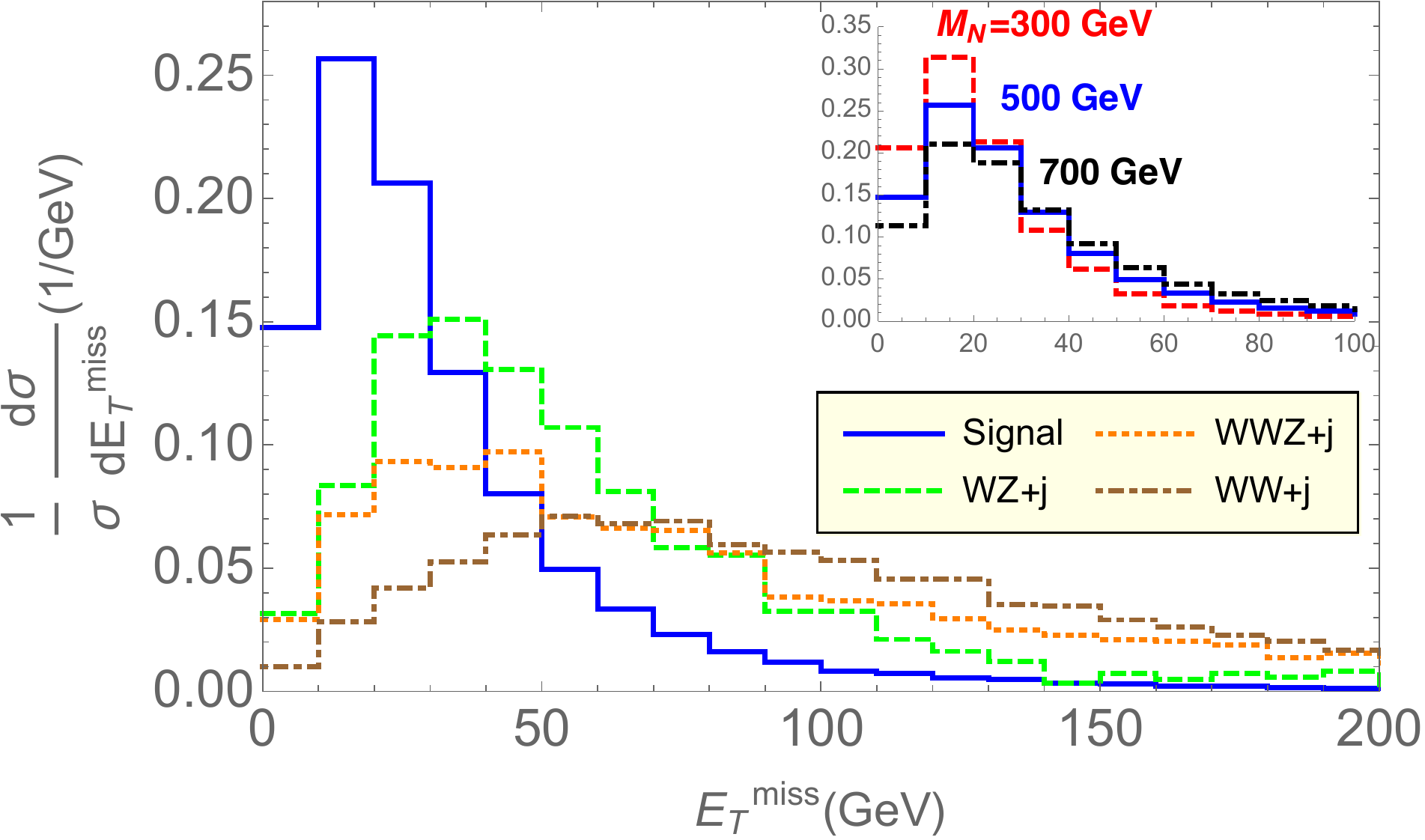}
\end{center}
\caption{ 
 Normalised differential distribution of events as a function of missing transverse momentum, after the application of the basic selection cuts; including $p^{J}_T > 100$ GeV. The distribution for $M_N=500$ GeV along with all dominant backgrounds is shown. Inset shows the variations for three benchmark signal points with $M_N = 300,\; 500$ and $700$ GeV.
}
\label{fig:dist_MET} 
\end{figure}

We implement the parton level event generation using {\tt MadGraph5-}{\tt aMC@NLO} \cite{MG, MG5, Alwall:2014hca} and signal model files are generated with {\tt FeynRules} \cite{Christensen:2008py, Alloul:2013bka}.  {\tt CTEQ6L} \cite{Pumplin:2002vw} is adopted for the parton distribution functions (PDF) and the factorization scale $\mu_F$ is set to the default MadGraph option. The showering, fragmentation and hadronization of the generated events were performed with {\tt PYTHIA6.4} \cite{Sjostrand:2006za}. 
The matching is done using the MLM scheme\cite{Hoche:2006ph}; based on a shower-kT algorithm with pT-ordered showers. For SM backgrounds, the matching scale {\tt QCUT} is set between 20 and 30 GeV. The showered events are passed through {\tt Delphes 3.3.2}~\cite{deFavereau:2013fsa} for detector level simulations with the default CMS card. The jets and associated substructure variables are constructed as described in section~\ref{sec:fat}.
\begin{figure}[t!]
\begin{center}
\includegraphics[bb=0 0 548 341,scale=0.6]{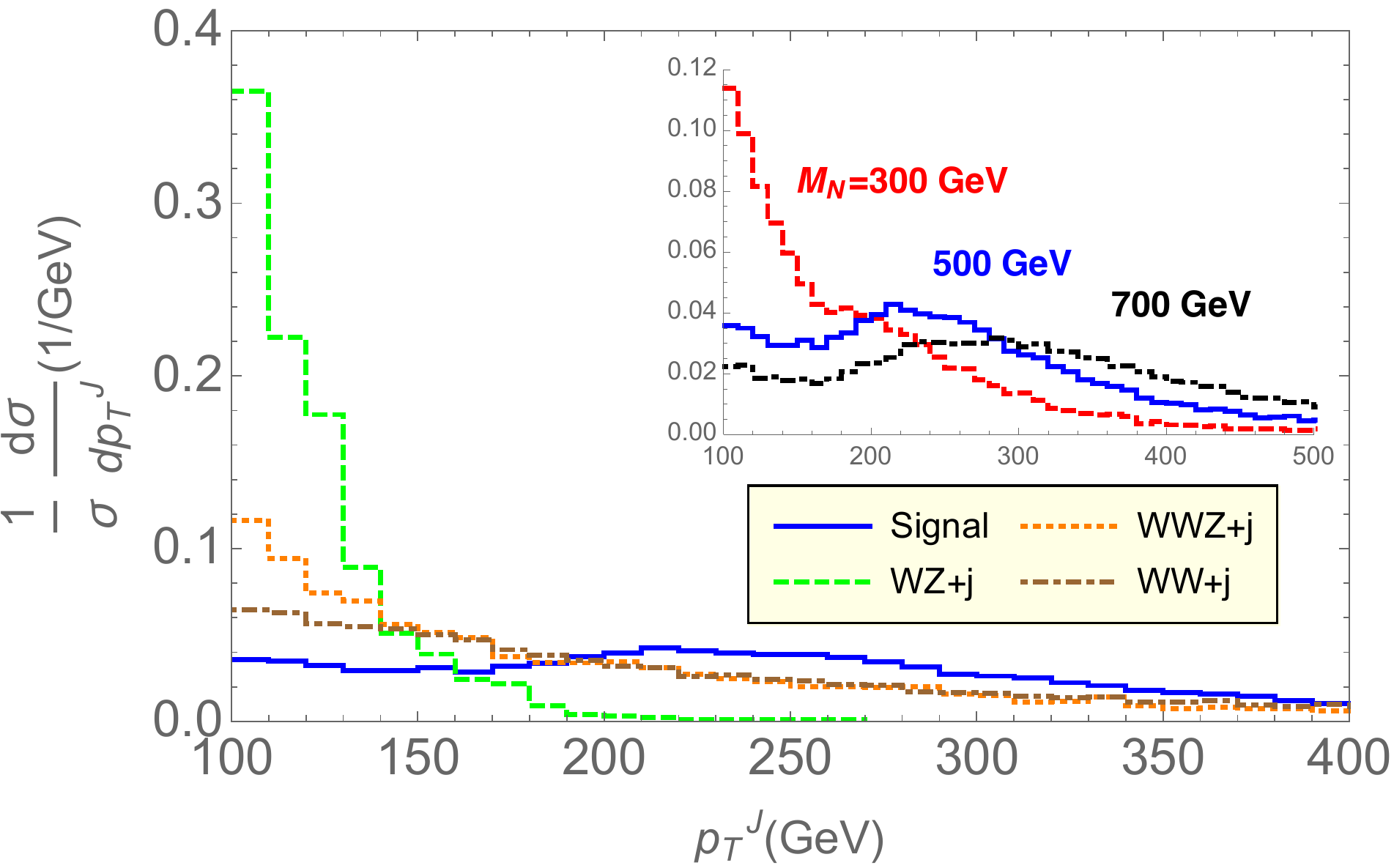} 
\end{center}
\caption{ 
Normalised differential distributions as a function of fat-jet (J) transverse momentum, again after the application of the basic selection cuts. $M_N=500$ GeV distribution with all dominant backgrounds are shown along with an inset showing the $M_N = 300,\; 500$ and $700$ GeV cases.
}
\label{fig:dist_PTJ} 
\end{figure}
 
 The $W^\pm W^\pm$ production cross-section is $\sigma_{W^\pm W^\pm}$ = 119.26 fb as calculated with the full next-to-leading order in perturbative QCD and electroweak corrections to the vector-boson scattering as well as its irreducible background \cite{Biedermann:2017bss}. For the $W^\pm Z$ and $W^\pm W^\pm Z$ channels cross-sections are 51.11 pb \cite{Grazzini:2016swo} and 197.41 fb \cite{Nhung:2013jta} in the NNLO and NLO QCD respectively. 
The effect of the next-to-leading order QCD correction for heavy neutrino production with arbitrary renormalization and factorization scale choices have been studied in \cite{Das:2016hof}. We consider corresponding cross-section for different heavy neutrino masses.
 
 To establish specific features and kinematic characteristics related to our RHN signal and backgrounds, we start by focusing on signal identification. Our prototypical signal is Same Flavour ($\mu^\pm$ ) SSDL, in association with a fat-jet. We adopt the following selection criteria
 
\begin{itemize}
\item Muons $\mu^\pm$ are identified with a minimum transverse momentum $p^\mu_T > 10$ GeV and rapidity range $|\eta^\mu| < 2.4$, with a maximum efficiency of $95\%$. Efficiency decreases for $p^\mu_T$ above  $1\,\mathrm{TeV}$.
\item Only events with reconstructed di-muons having same sign are selected for further analysis.
\item Hard jets having at least $p_T^j > 10 \,\mathrm{GeV}$ and $|\eta^j| < 2.4$  are identified.
\item Candidate fat-jets are to be identified, following the criteria in section~\ref{sec:fat} (an $R=0.8$, CA jet with $|\eta^J| < 2.4$).
\item We identify the hardest fat-jet with the $W^\pm$ candidate jet ($J$), and this is required to have $p^{J}_T > 100$ GeV.
\end{itemize}
 
 The above {\sf basic selection criteria}  are like primary level cuts required for effective signal identification. The last requirement is to ensure robust fat-jet properties.  As argued earlier, features of the boosted fat-jet are rather more prominent for large $M_N$; showing up emphatically for 300 GeV and above. In the next section we introduce some additional event criteria and then illustrate various results by considering several signal benchmark points.

\begin{figure}[t]
\begin{center}
\includegraphics[bb=0 0 548 341,scale=0.6]{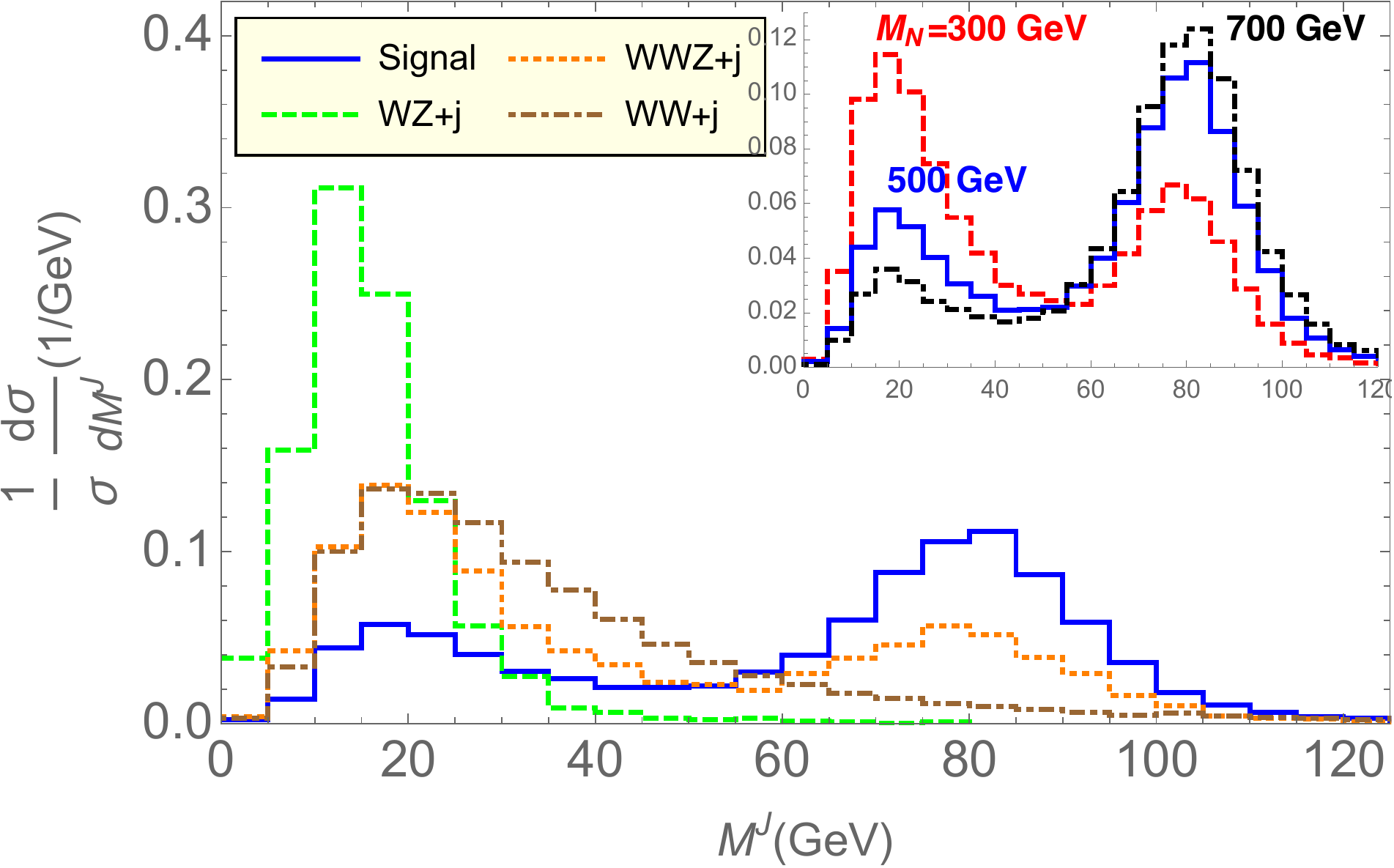}
\end{center}
\caption{ 
Normalised differential distributions as a function of fat-jet (J) invariant mass in case of same sign di-lepton + fat-jet production channel after the application of the basic selection cuts including $p^{J}_T > 100$ GeV. Distributions of one signal region with all dominant backgrounds are shown in the plot. Inset shows the variation for three benchmark signal points with $M_N = 300,\; 500$ and $700$ GeV.
}
\label{fig:dist_MJ} 
\end{figure}
%

\section{Results and Discussion}
\label{sec:res}

In the previous section,  basic selection criteria were set. We are now in a position to identify specific features and kinematic characteristics that can further differentiate RHN events from SM backgrounds. To highlight the differences, we focus on four key characteristic distributions, considering backgrounds along with three signal benchmark points (based on $M_N = 300,\; 500$ and $700$ GeV).

Figure~\ref{fig:dist_MET} illustrates the normalized differential distribution of events as a function of missing transverse momentum, after the application of the basic selection cuts. Missing transverse momentum (MET) is calculated from the contributions of isolated electrons, muons, photons and jets along with unclustered deposits. Our signal of interest from RHN involves no missing particle at the detector and is thus expected to have low MET. The only MET contributions may be from the mismeasurement of hard jets. On the contrary, in almost all relevant background processes, leptons originate from $W^\pm$ along with a neutrino. The neutrinos are not detected and contribute to a large MET. Distribution of one prototypical signal region with all dominant backgrounds is shown in the plot. It clearly shows the larger MET contribution for the backgrounds. Inset shows the same distribution for three benchmark signal points, $M_N = 300,\; 500$ and $700$ GeV. Distributions are very mildly sensitive to $M_N$, since heavier masses contribute to harder boosted jets and the jet-energy mis-measurements have a $P_T$ dependence.

\begin{figure}[t]
\begin{center}
\includegraphics[bb=0 0 548 336,scale=0.6]{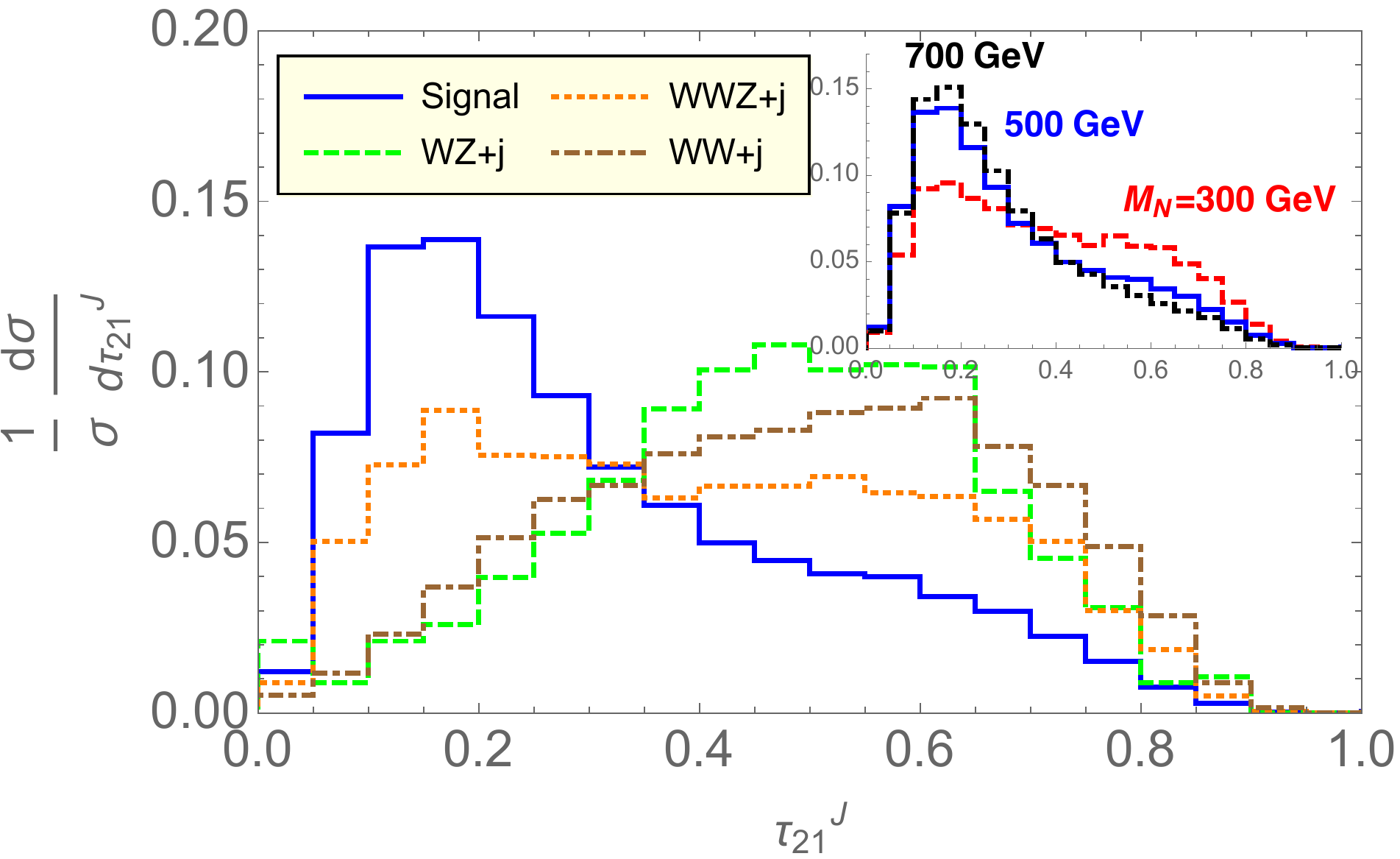}
\end{center}
\caption{ 
Normalised differential distributions as a function of two to one N-subjet ratio $\tau_{21}^J$ for fat-jet (J) in case of same sign di-lepton + fat-jet production channel after the application of the basic selection cuts including $p^{J}_T > 100$ GeV. Distributions of one signal region with all dominant backgrounds are shown in the plot. Inset shows the variation for three benchmark signal points with $M_N = 300,\; 500$ and $700$ GeV.
}
\label{fig:dist_tau21J} 
\end{figure}
%

Figure~\ref{fig:dist_PTJ} presents the normalized differential distributions for the fat-jet transverse momentum in a similar way. Here, minimum $P_T^J$ of 100 GeV has already been imposed. As we discussed in section~\ref{sec:fat}, $P_T^J$ is the vector sum of all constituent four momenta in $J$. Signal distribution is noticeably harder compared to background distributions, which fall faster. Comparison of different signal distributions is also quite interesting. As expected, heavier $M_N$ produces harder $J$ candidates.

Imposing a minimum $P_T^J$ selection brings out marked differences in the distribution of $M_J$ and $\tau_{21}^J$, between signal and backgrounds events. These jet shape variables can be very powerful in further containing the backgrounds. 
Two fat-jet invariant mass peaks are evident from the figure~\ref{fig:dist_MJ}. Second peak at around 80 GeV reflects the jet mass of $W$ like fat-jet $J$. This peak is absent for those backgrounds where fat-jet is faked by QCD jets. Only the triple gauge boson background, where fat-jet can originate from hadronic decay of one of the W's, provide some contamination to signal. The signal plots in the inset are also quite instructive, showing the significant $W$ like fat-jet contributions for higher $M_N$. The small spurious peak is due to events where some four-momenta from the hadronically decaying boosted-$W^\pm$ is missed in the jet clustering. This spurious peak around $20\,\mathrm{GeV}$ may be reduced by imposing a larger $P_T^J$. This would of course cut down the signal as well, and we find $P_T^J \gtrsim 100\,\mathrm{GeV}$ to provide the most optimal signal significance.

Another excellent discriminant to tag a hadronic two-pronged object is $\tau_{21}^J=\tau_2/\tau_1$, as we discussed in section~\ref{sec:fat}. Corresponding distributions are shown in figure~\ref{fig:dist_tau21J}.  $\tau_{21}^J$ for $W^\pm$-like fat-jets peak around small values and this is clearly visible in figure~\ref{fig:dist_tau21J}. It becomes more prominent for larger $M_N$, as the inset figure shows, due to the $J$ being more boosted.

It is important to reemphasize that the choice of a higher, minimum $P_T^J$ effectively selects purer, $W^\pm$-like fat-jet events, but probably at the cost of some signal. This is essentially reflected in the larger event fractions in the higher (lower) peaks for $m^J$ ($\tau_{21}^J$). This would result in a sharper peak and background reductions. We find $P_T^J > 150\,\mathrm{GeV}$ to be optimal for selecting events, while maintaining good signal significance, as mentioned.

We list below our final event selection criteria motivated by the kinematic distributions. 

 \begin{itemize}
\item Leading muon should have $p_T  (\mu_1) > 20$ GeV and the next hardest muon must have $p_T  (\mu_2) > 15$ GeV.
\item Minimum invariant mass for the same sign muon pair must satisfy $m_{\mu \mu} > 50$ GeV. This is easily satisfied for the signal events, and can control backgrounds with non-prompt muon pairs.
\item Lacking any missing particles for our signal, require $E_T^{\rm{miss}} < 35$ GeV. This can control background events with large MET contributions.
\item The hardest, reconstructed fat-jet must have $p_T^J > 150$ GeV.
\item We also demand the invariant mass of the hardest, reconstructed fat-jet to satisfy $M_{J} > 50$ GeV. In principle one may use a mass window around the $W^\pm$ mass, but we find that a simple lower bound suffices.
\item The N-subjettiness ratio corresponding to the reconstructed fat-jet must satisfy $\tau_{21}^J  < 0.5$.
\end{itemize}
With these we are able to achieve very significant background elimination, relative to the signal.

 \begin{table}[t]
 \tiny
 \centering
 \begin{tabular}{|l||c|c|c||c|c|c|}
 \hline
  Cut                                             & \multicolumn{3}{c||}{Signal for $M_N$}                                                                       & \multicolumn{3}{c|}{Background}          \\ \hline\hline
                                                     & $300$ GeV           & $500$ GeV           & $700$ GeV            & $WW$+j                   & $WZ$+j                  & $WWZ+j$               \\ \hline
 Pre-selection +                           & & & & & & \\
 $\mu^\pm \mu^\pm+ J$               &{82.2+ 45.2}           &{36.6+23.4}           &{19.2+13.0}               &{2717.5+2597.0}        &{9881.3+7639.3}     &{252.1+240.4}   \\ 
 $p^{J}_T > 100$ GeV                  &[100\%]                 &[100\%]                  &[100\%]                   &[100\%]                      &[100\%]                   &[100\%]               \\ \hline
 $p_T  (\mu_{1,2}), m_{\mu\mu}$ & 79.5+ 39.8           &33.02+ 20.3             & 15.6+9.2                 & 2255.7+2132.1   & 5496.6+5074.1        & 208.0+193.4        \\ 
                                                     &[94\%]                   &[88\%]                     &[77\%]                        &[83\%]                       &[60\%]                         &[82\%]                   \\ \hline
 $E_T^{\rm{miss}} < 35$ GeV       & 66.3+27.4            & 28.5 +18.1             & 10.0+7.6                  & 260.8+163.2         &189.9+188.1            & 24.2+ 19.6          \\
                                                      &[74\%]                   &[77\%]                     &[55\%]                        &[7.9\%]                       &[2.2\%]                         &[8.9\%]                   \\ \hline
 $p_T^J > 150$ GeV                     & 35.1+20.6              & 15.2+ 10.5             & 8.3+6.0                  & 152.4+91.4          &  36.5+ 27.2          &14.14+12.4           \\
                                                      &[44\%]                   &[58\%]                     &[44\%]                        &[4.5\%]                       &[0.4\%]                         &[5.3\%]                   \\ \hline
 $M_{J} > 50$ GeV                       &29.3+16.9              & 20.9+ 10.2             & 6.6+4.4                  & 34.0+26.6              &11.6+8.5                  & 6.6+5.0             \\
                                                      &[36\%]                   &[42\%]                     &[34\%]                        &[1.1\%]                       &[0.1\%]                         &[2.3\%]                   \\ \hline
 $\tau_{21}^J  < 0.5$                     &26.7+13.7              & 13.2+7.2                 & 5.4+2.8                  & 17.5+15.9                & 5.9+5.2                 & 3.0+2.8             \\
                                                      &[32\%]                   &[34\%]                     &[25\%]                        &[0.6\%]                       &[0.06\%]                         &[1.2\%]                   \\ \hline
 \end{tabular}
 \caption{The effectiveness of different variables in minimizing backgrounds is iilustrated in the form a cut flow. The two numbers correspond to expected events in $\mu^+ \mu^+$ and $\mu^- \mu^-$ channels. We adopt a typical mixing angle $|V_{\mu N}|=0.03$. The numbers are for an integrated luminosity of $3000$fb$^{-1}$, at the 13 TeV LHC.}
 \label{tab:cut-flow}
 \end{table}

Now we present our results. The effects of the different cuts, as we have motivated, are summarized in Table~\ref{tab:cut-flow} in the form of a cut-flow. Three reference RNH benchmark points are presented with masses $300$ GeV, $500$ GeV and $700$ GeV. It is quite clear, in reference to the different distributions shown earlier, that the choice of these cuts are extremely efficient in controlling the large SM backgrounds. This enables the RHN signal to be probed to a significant mass range, or alternatively to smaller mixing angles, at the LHC. 

\begin{figure}[t]
\begin{center}
\includegraphics[bb=0 0 360 339,scale=0.75]{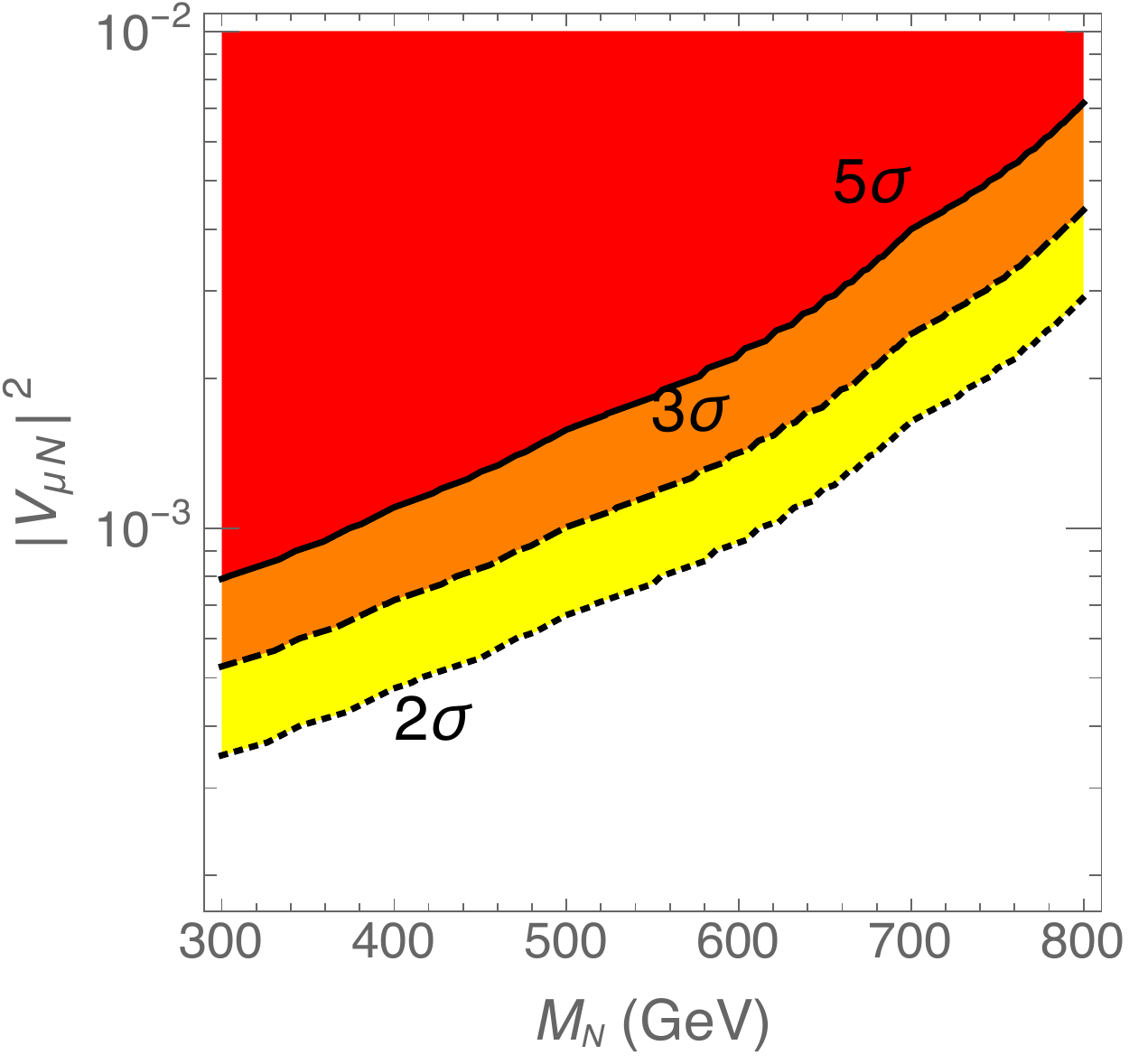}
\end{center}
\caption{
Exclusion limit in terms of heavy neutrino mass $M_N$ and  $|V_{\mu N}|^2$  at the 13 TeV LHC.}
 \label{fig:sig_con}
\end{figure}

The statistical significance ($\cal{S} $) of the observed signal events ($S$) over the total SM background events ($B$) has been calculated using
\bea
{\cal S} &=& S/ \sqrt{B}   \text{\hspace{158pt} for $5 \sigma$ significance,} \\
{\cal S} &=& \sqrt{2 \times \left[ (S+B)  \ln(1+\frac{S}{B}) - S \right] }  \text{ \hspace{25pt} for $2\sigma$ and $3 \sigma$ significance.}
\eea
Figure~\ref{fig:sig_con} displays the significance contours in the $(M_N, |V_{\mu N}|^2)$ plane.  These contours reflect the extensive capability of RHN searches augmented by jet substructure techniques. One obtains interesting limits all the way from $M_N=300$ GeV with $|V_{\mu N}|^2=3.4 \times 10^{-4}$  to $M_N=800$ GeV with $|V_{\mu N}|^2=2.9 \times 10^{-3}$. 
Additional production channels contributing to the heavy neutrino production such as $\gamma- W^\pm$ fusion is expected to increase the net cross-section at high energy collider and potentially improve the exclusion limits further especially for heavier masses  \cite{ Dev:2013wba, Alva:2014gxa}. In that sense our present estimates are conservative.

\begin{figure}[t]
\begin{center}
\includegraphics[bb=0 0 554 346,scale=.8]{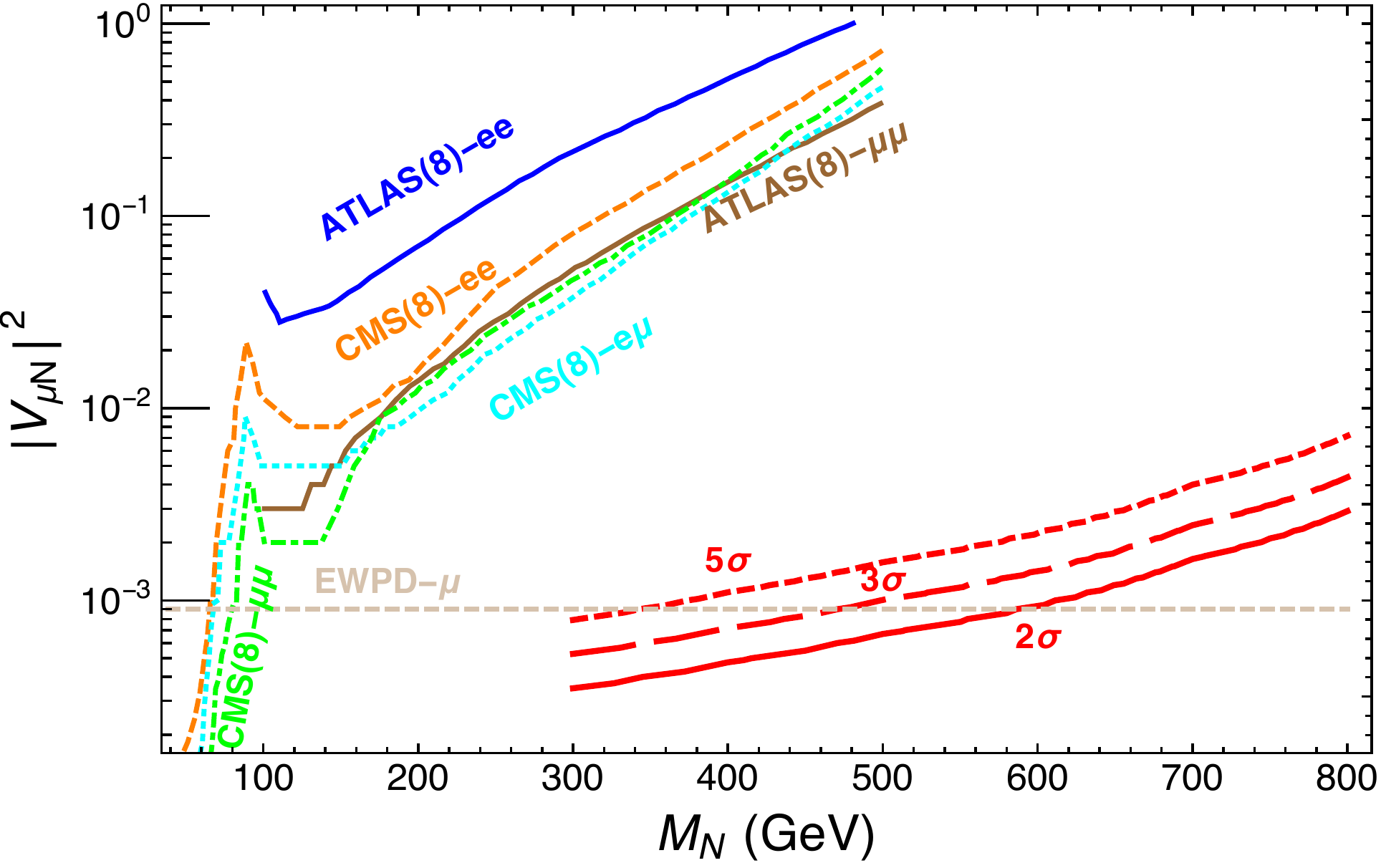} 
\end{center}
\caption{
Exclusion limit in terms of heavy neutrino mass $M_N$ and $|V_{\mu N}|^2$  at the 13 TeV LHC with other available limits.}
 \label{fig:sig}
\end{figure}

It is instructive to compare our projected collider limits with existing LHC limits, as well as indirect EWPD bounds. This is shown in figure~\ref{fig:sig}. There are currently no limits above $M_N=500$ GeV, from any experiment. From ATLAS searches \cite{Aad:2015xaa}  at $\sqrt{s}=8\,\mathrm{TeV}$, the blue solid line shows the limits for the $e^\pm e^\pm jj$ channel and the brown solid line for the $\mu^\pm \mu^\pm j j$ channel. The orange dashed line shows the limits for $e^\pm e^\pm j j$ and the green dot-dashed line shows limits for $\mu^\pm \mu^\pm j j$, both from CMS \cite{CMS8:2016olu, Khachatryan:2015gha}. The light gray solid line stands for the EWPD limit for $\mu$ flavor mixings \cite{Achard:2001qv, delAguila:2008pw, Akhmedov:2013hec}.

Note that our event selection criteria is optimized to the lower side of the heavy neutrino mass and we have applied the same cuts universally for the full signal mass range. Now, we have already observed in the inset plots of Figs.~\ref{fig:dist_MET}-\ref{fig:dist_tau21J}, that distributions change with $M_N$. Hence, there is sufficient room left to improve our results for higher $M_N$, by focused optimizations at each mass point. Instead of fine tuning the analysis, our main aim here was to demonstrate the efficacy and usefulness of jet substructure analysis in the general RHN collider search context \cite{Bhardwaj:2018lma}.

\section{Conclusions}
\label{sec:conc}
Seesaw models can naturally incorporate the existence of tiny neutrino masses and flavor mixings through simple extensions of the SM, many of which have Majorana RHNs. Such RHNs, if they exist at the TeV scale, can be produced and detected at the LHC. Searches have been performed for these states in the dilepton+dijet channel. Here we propose for the first time a search in the dilepton+fat-jet channel, leveraging jet substructure methods which can very significantly increase the LHC reach for these RHN states. In our case we considered the mass region $M_N \geq 300 $ GeV. We used the unique kinematic characteristics of a fat-jet -- such as $P_T^J$, $M^J$ and $\tau_{21}^J$-- to W-tag it and it was found that this helps greatly in discriminating the RHN SSDL+fat-jet signal from backgrounds. Exclusion limits are obtained by computing signal significance and the bounds we obtain are many orders of magnitude stronger than current LHC limits. 
 
\bigskip
\acknowledgments
We thank Akanksha Bhardwaj for independent validation of this work and also facilitating JaxoDraw \cite{Binosi:2003yf,Binosi:2008ig} generated diagram.


\end{document}